\documentclass[Crown
, times, doublespace,
sagev]{sagej}

\usepackage{moreverb,url}

\usepackage[colorlinks,bookmarksopen,bookmarksnumbered,citecolor=red,urlcolor=red]{hyperref}
\usepackage{multirow}

\newcommand\BibTeX{{\rmfamily B\kern-.05em \textsc{i\kern-.025em b}\kern-.08em
T\kern-.1667em\lower.7ex\hbox{E}\kern-.125emX}}

\def\volumeyear{2016}

\begin{document}
\runninghead{Sisti and Gutman}

\title{A Bayesian Method for Adverse Effects Estimation in Observational Studies with Truncation by Death}

\author{Anthony Sisti, PhD\affilnum{1}, Andrew Zullo\affilnum{2}, PhD and Roee Gutman, PhD \affilnum{1}}

\affiliation{\affilnum{1} Department of Biostatistics, Brown University
\ \\ \affilnum{2} Department of Epidemiology, Brown University
}

\corrauth{Anthony Sisti, Department of Biostatistics,
Brown University,
Providence, RI,
USA.}

\email{anthony\_sisti@brown.edu}

\begin{abstract}
Death among subjects is common in observational studies evaluating the causal effects of interventions among geriatric or severely ill patients. High mortality rates complicate the comparison of the prevalence of adverse events (AEs) between interventions. This problem is often referred to as outcome "truncation" by death. A possible solution is to estimate the survivor average causal effect (SACE), an estimand that evaluates the effects of interventions among those who would have survived under both treatment assignments. However, because the SACE does not include subjects who would have died under one or both arms, it does not consider the relationship between AEs and death. We propose a Bayesian method which imputes the unobserved mortality and AE outcomes for each participant under the intervention they did not receive. Using the imputed outcomes we define a composite ordinal outcome for each patient, combining the occurrence of death and the AE in an increasing scale of severity. This allows for the comparison of the effects of the interventions on death and the AE simultaneously among the entire sample. We implement the procedure to analyze the incidence of heart failure among geriatric patients being treated for Type II diabetes with sulfonylureas or dipeptidyl peptidase-4 inhibitors.
\end{abstract}


\maketitle

\section{Introduction}
\label{sec:intro}
In the United States in 2023, about 1 in 10 individuals had diabetes, with approximately 90-95\% of these cases classified as type 2 diabetes mellitus (T2DM) \citep{CDCdiabetes}. The disease is even more prevalent among older adults \citep{kirkman2012diabetes}. Sulfonylureas (SUs) are among the most common stand alone treatments
for nursing home (NH) residents diagnosed with T2DM; however, usage rates of  dipeptidyl peptidase-4 inhibitors (DPP4Is)  have increased in recent years \citep{ Diabetes_info, Diabetes_drug_use}. Studies among the general population have been inconclusive on the relative safety and efficacy of SUs and DPP4Is. Some studies have shown that SUs are associated with increased cardiovascular and hypoglycemia risk, while DPP4Is are associated with an increased risk of heart failure related hospitalizations \citep{SU_v_DPP4I, DPP4I_heartfailure_hosp, DPP4I_HF_effect}. DPP4Is have also been found to be more effective for glycemic control compared to SU \citep{SU_v_DPP4I, fadini2018_compareSU_DPP4I, DPP4I_HF_effect}. Most of these studies have not been conducted among NH residents who are disproportionately effected by T2DM and are susceptible to negative side effects of medications \citep{zullo_compareSU, zullo2022effects, Zullo_DiabetesTreatments,Diabetes_info}. 

NH residents are often excluded from randomized studies because of efficacy concerns, ethical considerations and practical challenges of conducting clinical trials in nursing homes \citep{Old_people_clinical}. In addition, because adverse events are generally rare, randomized studies with adverse events as the primary outcomes require large sample sizes to identify significant effects. Observational studies can address these limitation by relying on larger samples that include populations not commonly recruited to clinical trials. However, in observational studies, the decisions on which patients receive the interventions are often confounded with the outcomes.

One observational study that compared SUs and DPP4Is among older adults was a national retrospective cohort study of long-stay NH residents  \citep{zullo_compareSU}. The study estimated the effects of DPP4Is versus SUs on severe adverse glycemic events, cardiovascular events, and death. The study concluded that DPP4I users had similar incidence of cardiovascular events and death, and a lower incidence of hypoglycemic events, over a 1-year time frame compared to SU users. However, many patients died before the 1-year follow up time was complete.  This study considered death and the adverse event separately, which does not allow for a joint evaluation of the relationships between DPP4Is and SUs, adverse events and deaths. It also ignores the fact that a person is no longer at risk for an adverse event after death. \\
\indent Estimating the effects of interventions on the occurrence of adverse events in the presence of mortality complicates the analysis because the occurrence of an adverse event at follow-up is not defined for individuals who die before the end of the follow up period \citep{MethReview,Probs, rubinSACE, RubZhangSACE}. One approach to address ``truncation" by death is to estimate the intention to treat effect (ITT), which considers the difference in proportions of the adverse event that occurred between the two study arms. This approach utilizes the entire sample, but differences in mortality across interventions may result in misleading conclusions.

A different approach to address truncation by death relies on the principal stratification framework to estimate the survivor-average causal effect (SACE)  \citep{ OGprinStrat, rubinSACE, RubZhangSACE}. This approach stratifies participants by mortality status under both interventions, and estimates the effect of the intervention on the outcome among participants who would have survived under both interventions. Many applications that estimate SACE rely on the monotonicity  and stochastic dominance assumptions \citep{rubinSACE}. The monotonicity assumption states that survival under the active intervention is the same or better than survival under the control intervention. However, in studies that compare two active interventions, this assumption may not be plausible. The stochastic dominance assumption requires that, on average, those who would survive under both interventions are less likely to experience the adverse event than those who would have died under either intervention. This assumption may also be violated in studies comparing two active interventions, because longer exposure to an intervention may lead to higher probability of experiencing an adverse event. Several approaches that modify these assumptions have been proposed to estimate the SACE in randomized and observational studies \citep{ExtraSACE5, ExtraSACE2,  ExtraSACE3}, but procedures that use the SACE ignore individuals who would have died under either of the treatments. This includes individuals who experienced an adverse event that leads to death under one of the interventions. As such, the SACE may not provide a complete toxicity profile for the two interventions.
\\
\indent Another approach to address the truncation problem is to define a composite outcome that combines death and the adverse event \citep{RCTReview}. This is commonly implemented by defining a binary outcome that is equal to 1 if either death or the adverse event occur, and zero otherwise. This composite outcome is well defined for all participants in the study, and common methods for estimating the treatment effects with binary outcomes can be applied in randomized  \citep{CompositeTrials, CompositeTrials2} and observational studies \citep{CompositeObservational,CompositeObservational2, RoeePaper}. Interpretation of the composite approach is practically relevant when the adverse event and mortality have similar utility for all patients, and when the interventions influence both mortality and the adverse event in the same direction. However, when the rate of either the adverse event or mortality is high for one of the interventions, it is difficult to assess differences between the interventions on the less prevalent outcome \citep{MethReview, BadComp}. \\
\indent Another solution is to define a composite ordinal outcome that combines the occurrence of death and the adverse event in an increasing scale of severity.
In randomized clinical trials, the desirability of outcome ranking (DOOR) was proposed as a possible ordinal outcome that combines the occurrence of multiple outcomes and ranks these combinations by their desirability \citep{DOOR}. The DOOR distributions are compared between the study arms by estimating the probability that an individual selected at random will have a better DOOR if assigned to the active intervention. Non-parametric tests and interval estimates can be used to evaluate whether this probability exceeds 50\%.  Analysis based on DOOR has been restricted to randomized studies and to compare marginal distributions without adjustments for covariates. Another composite statistic that has been proposed for randomized controlled trials is the win ratio \citep{WinRatio}. Among matched pairs of units, one receiving the active intervention and one receiving the control intervention, the win ratio estimates how often a unit under the intervention fared better than a unit under the control. A "winner" is assigned in each matched pair based on which unit had the more desirable outcome first, or lasted longer without experiencing an adverse outcome. The number of ``winners" under each intervention is computed and used to calculate the win ratio. The win ratio does not adjust for covariates and it relies on time to event data. In addition, when risk-matching among patients is not possible, obtaining confidence intervals and $p$-values for the win ratio is complex \citep{WinRatio}.

\indent We propose a new Bayesian method for observational studies that fits two conditional models for the adverse event and death in each study arm. Based on these models, we multiply impute the unobserved outcomes for each participant under the opposite intervention and construct a composite ordinal outcome. The proposed method generates statistically valid point and interval estimates for any causal estimand with ordinal outcomes. We apply this method to compare the 180-day risk of death and heart failure following the  initiation of a SU or DPP4I among NH residents with T2DM.  We show that  a randomly selected patient is estimated to have a 5\% higher risk of having a worse outcome under DPP4I than of having a worse outcome under SU, but this result was not significant at the 5\% nominal level.

\section{Methods}
\label{methodSec}
\subsection{Notation and Definitions}

In a sample from a population of $N$ units indexed $i=1,..,n\leq N$, let $n_0$ be the number of units receiving the control intervention and $n_1 = n - n_0$ units receiving the active intervention, where the control intervention may represent another active intervention. Let $W_{i} \in \{0,1\}$ be an indicator that is equal to 0 if unit $i$ received the control intervention, and 1 if it received the active intervention. We define $\mathbf{Y}_i(0) = \{A_i(0),D_i(0)\}$ to be a joint outcome indicating whether an adverse event, $A_i(0)$, or death, $D_i(0)$,  occurred between the initiation of the control intervention and the end of the study period for unit $i$. Similarly, let $\mathbf{Y}_i(1)=\{A_i(1),D_i(1)\}$ be the joint outcome for the active intervention. We write $\mathbf{Y}(0)= \{\mathbf{Y}_i(0)\}_{i=1}^N$ and $\mathbf{Y}(1)=\{\mathbf{Y}_i(1)\}_{i=1}^N$ to denote the collection of joint outcomes for the sample. Because individuals can receive only one of the interventions at a specific time point, only one of $\mathbf{Y}_i(0)$ and $\mathbf{Y}_i(1)$ can be realized and observed \citep{role_randomization}. Assuming the stable unit treatment value assumption (SUTVA) \citep{RubinSutva}, the observed and missing outcomes are, 
\vspace{-0.5pt}
\begin{align*}
    \mathbf{Y}_{i}^{obs}&=\mathbf{Y}_{i}(1)W_{i}+\mathbf{Y}_{i}(0)(1-W_{i})\\\mathbf{Y}_{i}^{mis}&=\mathbf{Y}_{i}(1)(1-W_{i})+\mathbf{Y}_{i}(0)(W_{i}).
\end{align*}

\noindent  For each unit, we also observe a set of $P$ covariates, $\mathbf{X}_i = \{X_{i}^1, ... , X_{i}^P\} $,  that are recorded prior to initiation of the intervention. \ \\
\indent To estimate the causal effects of the intervention, we need to identify the probability that units received the active intervention, $P(\mathbf{W}|\mathbf{Y}(1),\mathbf{Y}(0),\mathbf{X})$, commonly referred to as the assignment mechanism  \citep{role_randomization}. Assuming that the assignment mechanism is strongly ignorable \citep{Propensity}, and that the units are independent, we have
\[
P(\mathbf{W}|\mathbf{Y}(0),\mathbf{Y}(1),\mathbf{X}) = \prod_i^N P(W_i=1|\mathbf{Y}_i(1), \mathbf{Y}_i(0), \mathbf{X}_i, \phi) = \prod_i^N P(W_i=1| \mathbf{X}_i, \phi) = \prod_i^N e(\mathbf{X}_i), \]

\noindent where $\phi$ comprises the parameters governing this distribution, and $e(\mathbf{X}_i)$ is the propensity score for unit $i$ \citep{PropensityScore}. In randomized controlled trials,  the propensity score is known as part of the design phase. In observational studies, the probability of receiving each of the interventions is unknown, and only an estimate of it, $\hat{e}(\mathbf{X}_i)$, is available. Multiple procedures have been described for estimating the propensity score \citep{westreich2010propensity, stuart2010,guo2020propensity}. We assume that the propensity scores have been estimated using any procedure selected by the investigators in the design phase of the study. Our method will utilize the estimated propensity score at the analysis stage.

\subsection{Composite Ordinal Outcome and Estimands} \label{estimands}

Estimands, which are functions of unit-level potential-outcomes, are used to summarize the effects of an intervention across a population of interest \citep{RoeePaper, role_randomization}. Possible estimands that are defined for the entire population and consider death and the adverse event outcome separately include the difference in proportions, $Pr(A_i(1) = 1)-Pr(A_i(0)=1)$ and $Pr(D_i(1) = 1)-Pr(D_i(0)=1)$, and their corresponding risk and odds ratios. These estimands do not consider the correlations between the outcomes, which may lead to misleading conclusions when mortality is differentiable between the two treatments. Other possible estimands to address this issue are based on a binary composite outcome. A binary composite outcome is equal to 1 if either of the events occur and zero otherwise. For example, the difference in proportions estimand for this outcome is $Pr(A_i(1)=1\cup D_i(1) = 1)-Pr(A_i(0)=1 \cup D_i(0) = 1)$. This estimand may present an incomplete picture when the probability of mortality is much larger than the probability of adverse event and vice versa. A different estimand that attempts to address this limitation is the SACE. The SACE estimates the effects of the intervention on the adverse event among the sub-population that survives under both interventions: $Pr(A_i(1)=1|D_i(0)=D_i(1)=0)-Pr(A_i(0)=1|D_i(0)=D_i(1)=0)$. This estimand may not present the entire toxicity profile of intervention, when mortality is higher for one of the interventions.

 To address these limitations we define a composite ordinal potential outcome $G_i(w)$ that combines deaths and adverse events in an increasing scale of severity: 
 \[
G_{i}(w) = \begin{cases}
1 & \text{if\ }\mathbf{Y}_{i}(w)=\left\{ 0,0\right\} \\
2 & \text{if\ }\mathbf{Y}_{i}(w)=\left\{ 1,0\right\} \\
3 & \text{if\ }\mathbf{Y}_{i}(w)=\left\{ 0,1\right\} \\
4 & \text{if\ }\mathbf{Y}_{i}(w)=\left\{ 1,1\right\} 
\end{cases} .
\]

\noindent This definition assumes that death is worse than undergoing an adverse event, but different rankings of severity and additional ordinal levels can be used in other situations. This outcome is well-defined on the whole sample, and enables researches to consider the associations between adverse events and death.  \\
\indent Let $p_k(w) = Pr(G_i(w)=k)$ for $k=1,...,4$; possible estimands for the composite ordinal outcome are $p_k(1)-p_k(0), \frac{p_k(1)}{p_k(0)}$ and $\frac{p_{k}(1)/(1-p_k(1))}{p_{k}(0)/(1-p_k(0))}$, which describe the difference in probabilities, the relative risk and the odds ratio of being in level $k$ of the ordinal outcome under the active intervention and the control intervention, respectively.  

A different estimand is  the distributional estimand for ordinal outcomes \citep{distcaus},

\begin{equation} \label{distributionalEstimand}
    \Delta_{j}=Pr\left( G_{i}(1)\leq j\right) -Pr\left( G_{i}(0)\leq j\right) =\sum_{l}\sum_{k\geq j}p_{kl}-\sum_{k}\sum_{l\geq j}p_{kl}\ \ \ 1\leq j\leq4,
\end{equation}

\noindent  where $p_{kl} = Pr(G_i(1) = k, G_i(0)=l) \text{ for }  k,l\in \{1,2,3,4\}$. This estimand is comprised of four values corresponding to each level of the ordinal outcome, with positive $\Delta_{j}$ indicating that an individual is more likely to experience level $j$ or lower under the active intervention than the control. 

Two other estimands were proposed by Lu et. al.\cite{lu_treatment_2018},
\[\tau_{10}=Pr\left( G_{i}(1)\geq G_{i}(0)\right) =\underset{k\geq l}{\sum\sum}p_{kl},\ \ \text{and} \ \ \kappa_{10}=Pr\left( G_{i}(1)>G_{i}(0)\right) =\underset{k>l}{\sum\sum}p_{kl}.\]

\noindent Estimand $\tau_{10}$ describes the probability that an individual has an outcome under the active intervention that is greater than or equal to their outcome under the control intervention, while $\kappa_{10}$ describes the same quantity but for the strictly greater case. The estimands $\tau_{01}$ and $\kappa_{01}$ correspond to $\tau_{10}$ and $\kappa_{10}$, but with the ordinal outcome being larger under the control intervention than the active intervention.  The estimands  $\kappa_{10}-\kappa_{01}$ and $\frac{\kappa_{10}}{\kappa_{01}}$ describe the difference in probabilities and relative risk of an individual doing strictly worse under the active intervention compared to doing strictly worse under the control intervention, respectively. The estimands $\tau_{10}$ and $\kappa_{10}$  can be used to define the Mann-Whitney estimand \citep{agresti_analysis_2010, BossRidit, AnalyCompEnd},

\[ U_{10}=Pr\left( G_{i}(1)>G_{i}(0)\right) + \frac{1}{2}Pr\left( G_{i}(1)=G_{i}(0)\right) =\frac{\kappa_{10}+\tau_{10}}{2}.\]

\noindent This estimand describes the probability that an individual selected at random has inferior outcomes under the active intervention compared to the control intervention, and $\frac{U_{10}}{1-U_{10}}$ describes the odds of this scenario. 

Let, $\pi_{kl}(w)=Pr\left( G_{i}(1-w)=l|G_{i}(w)=k\right) $, the conditional probability matrix under intervention $w$ is another possible estimand,

$$
\Pi(w)=\left(\begin{array}{cccc}
\pi_{11}(w) & \pi_{12}(w) & \pi_{13}(w) & \pi_{14}(w)\\
\pi_{21}(w) & \pi_{22}(w) & \pi_{23}(w) & \pi_{24}(w)\\
\pi_{31}(w) & \pi_{32}(w) & \pi_{33}(w) & \pi_{34}(w)\\
\pi_{41}(w) & \pi_{42}(w) & \pi_{43}(w) & \pi_{44}(w)
\end{array}\right)
$$

\noindent This matrix describes the probabilities that an individual would have outcome $l$ under intervention $1-w$ given that they have outcome $k$ under intervention $w$. Individual probabilities within the matrices can be subtracted, $\Pi(1)- \Pi(0)$ or divided $\Pi(1)/\Pi(0)$ to obtain risk differences or risk ratios corresponding to specific events, respectively.

\subsection{Bayesian Imputation of Counterfactual Outcomes} \label{sec:impute}

Deriving interval estimates for the estimands in Section 2.2 can be analytically complex, and Lu et al.\cite{lu_treatment_2018} describe sharp bounds to estimate $\tau_{10}$, $\kappa_{10}$ and $U_{10}$.  We view causal inference from a missing data perspective \citep{MissingData}, and extend a method introduced by Gutman and Rubin \cite{RoeePaperDich, RoeePaper} to multiply impute the unobserved potential outcomes. Let $\gamma = \nu(\mathbf{Y}(1), \mathbf{Y}(0))$, be a predefined estimand. A Bayesian approach to estimating $\gamma$ will condition on the observed data $\mathbf{Y}^{obs}, \mathbf{W}, \text{and} \  \mathbf{X}$ to estimate
\vspace{-0.5pt}
\begin{align*}
f(\gamma|\mathbf{Y}^{obs},\mathbf{X},\mathbf{W})&=\int f(\gamma|\mathbf{Y}^{obs},\mathbf{Y}^{mis},\mathbf{X},\mathbf{W})f(\mathbf{Y}^{mis}|\mathbf{Y}^{obs},\mathbf{X},\mathbf{W})d\mathbf{Y}^{mis}\\&=\int f(\gamma|\mathbf{Y}^{obs},\mathbf{Y}^{mis},\mathbf{X},\mathbf{W})f(\mathbf{Y}^{mis}|\mathbf{Y}^{obs},\mathbf{X})d\mathbf{Y}^{mis}.
\end{align*}

\noindent Because the distribution of $\gamma$ is a function of the potential outcomes, $\mathbf{X}$, and $\mathbf{W}$, only the conditional distribution of the missing outcomes, $f(\mathbf{Y}^{mis}|\mathbf{Y}^{obs},\mathbf{X})$, is required to estimate the distribution of $\gamma$ \citep{RoeePaper}. This distribution can be written as the product of the conditional distribution of the adverse event given the observed values, and the conditional distribution of death given the observed values and the adverse event, 
\vspace{-0.5pt}
\begin{align}\label{Dists}
    f(\mathbf{Y}^{mis}|\mathbf{Y}^{obs},\mathbf{X})&=f(\mathbf{D}^{mis},\mathbf{A}^{mis}|\mathbf{D}^{obs},\mathbf{A}^{obs},\mathbf{X}) \nonumber \\&=f(\mathbf{D}^{mis}|\mathbf{A}^{mis},\mathbf{D}^{obs},\mathbf{A}^{obs},\mathbf{X})f(\mathbf{A}^{mis}|\mathbf{D}^{obs},\mathbf{A}^{obs},\mathbf{X}).
\end{align}

 In observational studies with an unconfounded assignment mechanism, Gutman \& Rubin (2015) proposed to estimate the response surface, $E_{\theta_w}(Y(w)|X, \theta_w) = h_w(X,\theta_w)$, using a spline along the propensity score and linear adjustments for the other covariates. This method relies on subclasses of the propensity scores to balance the covariates across treatment groups while increasing efficiency using splines. In addition, it relies on linear adjustments for the components of the covariates orthogonal to the propensity score to gain additional efficiency and control for minor residual imbalances. Formally, we approximate the conditional distributions in Equation (\ref{Dists}) as

\begin{equation}\label{Yfunc}
 \tilde{Pr}(A_i(w) = 1 | \mathbf{X}_i, \theta^a_w) = g^{-1}(f^a_w(g(\hat{e}(\mathbf{X}_i)), \mathbf{B}^a_w) + \mathbf{X}_{i}^{*}\beta^a_w)   
\end{equation}
\noindent and,
\begin{equation} \label{Zfunc}
    \tilde{Pr}(D_i(w) = 1 | \mathbf{X}_i, \theta^d_w, A_i(w)) = g^{-1}(f^d_w(g(\hat{e}(\mathbf{X}_i)), \mathbf{B}^d_w) + \mathbf{X}^{*}_i\beta^d_w + \eta_w A_i(w)),
\end{equation} 

\noindent where $g$ is the logit function, $f_{w}^{a}$ and $f_{w}^{d}$ are splines over the propensity scores, $\theta^a_w = \{\mathbf{B}^a_w, \beta^a_w\}$ and  $\theta^d_w = \{\mathbf{B}^d_w, \beta^d_w, \eta_w\}$ are the sets of unknown parameters in Equations (\ref{Yfunc}) and (\ref{Zfunc}), respectively, and $\mathbf{X}_i^{*}=(X_{i}^1,...,X_{i}^{P-1})$, with $X_{iP}$ being omitted. 

Assuming that $\mathbf{Y}_i(1)$ and $\mathbf{Y}_i(0)$ are conditionally independent given $\mathbf{X}_i$, $\theta^{a}_w$ and $\theta^{d}_{w}$, and that the prior distributions of $\theta^{a}_w$ and $\theta^{d}_{w}$ are independent, the posterior distributions of $\{\theta^{a}_1,\theta^{d}_{1}\}$ and $\{\theta^{a}_0,\theta^{d}_{0}\}$ are independent. This assumption is formally known as no contamination of imputation across treatments \citep{EpiRubin}, and is made by many causal inference methods \citep{RoeePaper}. In addition, we assume that $P(\theta_w^a,\theta_w^d) \propto P(\theta_w^a)P(\theta_w^d)$, then $\theta_1^a$ and $\theta_1^d$, and $\theta_0^a$ and $\theta_0^d$, have independent posterior distributions. Let $\psi_{w}^a(\theta_w^a|\mathbf{D}^{obs},\mathbf{A}^{obs},\mathbf{X})$ and $\psi_{w}^d(\theta_w^d|\mathbf{D}^{obs},\mathbf{A}^{mis},\mathbf{A}^{obs},\mathbf{X})$ be the posterior densities of the parameters in equations (\ref{Yfunc}) and (\ref{Zfunc}) under treatment $w$ for adverse events and death,  respectively. To obtain $m=1,\ldots,M$ samples of $\mathbf{Y}(0)^{mis}$, we sample at iteration $m$, $\theta_0^{a(m)}$ and $\theta_0^{d(m)}$ from $\psi^a_0$ and $\psi_0^d$, respectively. Given $\theta_{0}^{a(m)}$ and $\theta_{0}^{d(m)}$, we sample $\mathbf{Y}_i(0)^{mis(m)}$ from the posterior predictive distribution implied by Equations (\ref{Yfunc}) and (\ref{Zfunc}). Similarly, $\mathbf{Y}_{i}(1)^{mis}$ can be sampled from the corresponding posterior predictive distributions.

When estimating the posterior distribution of $\gamma$, it is important to distinguish between finite-sample and super-population estimands, denoted $\gamma^{fp}$ and $\gamma^{sp}$, respectively. To estimate $\gamma^{sp}$ we need to consider the distribution of the covariates in the super-population, $F(\mathbf{X}_i)$ , which is not necessary when estimating $\gamma^{fp}$. Estimating $F(\mathbf{X}_i)$ is part of the design phase of observational studies because it does not involve any outcomes. When $\mathbf{X}$ is assumed to be a simple random sample from the population,  the empirical distribution, $\hat{\mathbb{F}}_{\mathbf{X}_i}$, can be used to approximate the super-population distribution of the covariates. Under other sampling mechanisms, different approaches, such as weighting, may be required to estimate $F(\mathbf{X}_{i})$. We now summarize the procedure for estimating $\gamma$:

\begin{enumerate}
    \item Estimate the propensity score, $\hat{e}(X_i)$ for each of the $n$ units in the sample, and partition these $n$ units into $K$ sub classes  based on the quantiles of the estimated propensity scores. 
    
    \item Based on equations (\ref{Yfunc}) and (\ref{Zfunc}) estimate $\tilde{Pr}(A_i(w) = 1 | \mathbf{X}_i, \theta^a_w)$,  and \\ $\tilde{Pr}(D_i(w) = 1 | A_i(w), \mathbf{X}_i, \theta^d_w)$, respectively, within treatment groups.
    
    \item Obtain $M$ samples of $\theta^a_0, \theta^d_0$ and $\theta^a_1, \theta^d_1$ from their corresponding posterior distributions.
    
    \item Using the $m^{\text{th}}$ sample, $\tilde{\theta}^{a(m)}_0$, sample $A_i(0)$ from a Bernoulli distribution with probability $\tilde{Pr}(A_i(0) = 1 | \mathbf{X}_i, \theta^{a(m)}_{0})$. Using $\tilde{\theta}_1^{a(m)}$ sample $A_i(1)$ for units with $W_i=0$ from a Bernoulli distribution with probability  $\tilde{Pr}(A_i(1) = 1 | \mathbf{X}_i, \theta^{a(m)}_{1})$.
    
    \item Using $\tilde{\theta}^{d(m)}_0$ and the sampled unobserved outcomes $\hat{\mathbf{A}}^{mis}$ from the previous step, impute the unobserved mortality for units with $W_{i}=1$ by sampling from a Bernoulli distribution with probability $\tilde{Pr}(D_i(0) = 1 | \mathbf{X}_i, \theta^{d(m)}_{0}, \hat{A}_i^{mis})$. Using $\tilde{\theta}_1^{d(m)}$, sample the unobserved death outcomes for units with $W_{i}=0$ from a Bernoulli distribution with probability  $\tilde{Pr}(D_i(1) = 1 | \mathbf{X}_i, \theta^{d(m)}_{1}, \hat{A}_i^{mis})$.
    
    \item Repeat steps 4 and 5 for $m=1,\ldots,M$.
    
    \item  For $m=1,\ldots,M$, let $\hat{\gamma}^{(m)} = \nu(\hat{\mathbf{Y}}(1)^{(m)}, \hat{\mathbf{Y}}(0)^{(m)})$ with $\hat{\mathbf{Y}_i}(w)^{(m)}=\{\hat{A}_i(w)^{(m)},\hat{D}_i(w)^{(m)}\}$ where
 \[
\hat{A_{i}}(w)^{(m)}=\begin{cases}
A_{i}^{obs} & \text{if\ }W_{i}=w\\
\hat{A}_{i}^{mis(m)} & \text{if\ }W_{i}\ne w
\end{cases}\ \ \ \ \ \ \ \ \ \ \ \ \hat{D_{i}}(w)^{(m)}=\begin{cases}
D_{i}^{obs} & \text{if\ }W_{i}=w\\
\hat{D}_{i}^{mis(m)} & \text{if\ }W_{i}\ne w
\end{cases}
    \]   
    \item  The point estimate of $\gamma$ is $\hat{\gamma}=\frac{1}{M}\sum_{i=1}^M \hat{\gamma}^{(m)}$. Let the sampling variance of $\hat{\gamma}^{(m)}$ be $\hat{U}^{(m)}$ ($\hat{U}^{(m)}$ = 0 for finite-sample estimands). We let $\bar{U}=\frac{1}{M}\sum_{i=1}^{M}\hat{U}^{(m)}$ define the within imputation variance and $B$ = $\frac{1}{M-1}\sum_{i=1}^{M}(\hat{\gamma}^{(m)}-\hat{\gamma})^{2}$ define the between imputation variance. The total estimate of the sampling variance for $\hat{\gamma}$ is $T= \bar{U} +(1+\frac{1}{M})B$.
    
    \item Posterior inferences are based on a $t$-distribution with $\nu_M$ degrees of freedom where ${\nu_{M}=(M-1)\left(\frac{T}{\left(1+M^{-1}\right)B}\right)^2}$ \cite{rubin2004multiple} . In small data sets, use the $t$-approximation derived by Barnard and Rubin \cite{Tapprox} and also described by Yuan\cite{yuan2010multiple} to estimate posterior intervals (Appendix Section \ref{CombinationRules}).

    
\end{enumerate}

\noindent When $M$ is large enough, posterior interval estimation for finite-sample estimands can also be derived using the percentiles of the distribution of $\gamma^{(m)}$ where $m =1,...,M$.  Some super-population estimands can be estimated using $\theta_{0}^{a(m)},\theta_{0}^{d(m)},\theta_{1}^{a(m)},\theta_{1}^{d(m)}$, and omitting steps 4-9 in the estimation procedure. For example, the super-population average treatment effect on adverse events, $\gamma^{sp}=\gamma(\theta_{1}^{a},\theta_{0}^{a})=E_{\mathbf{X}_{i}}[\gamma(\theta_{1}^{a},\theta_{0}^{a},\mathbf{X}_{i})|\theta_{1}^{a},\theta_{0}^{a}]$, where
\begin{align*}
\gamma(\theta_{1}^{a},\theta_{0}^{a},\mathbf{X}_{i})&=E_{A_{i}(1),A_{i}(0)}[A_{i}(1)-A_{i}(0)|\theta_{1}^{a},\theta_{0}^{a},\mathbf{X}_{i}]\\&=\tilde{Pr}(A_{i}(1)=1|\mathbf{X}_{i},\theta_{1}^{a})-\tilde{Pr}(A_{i}(0)=1|\mathbf{X}_{i},\theta_{0}^{a}).
\end{align*}

\noindent Assuming that the data originates from a simple random sample, and using the empirical distribution to approximate the covariate distribution in the super-population, we have
\[\gamma^{sp(m)}=\frac{1}{n}\sum_{i=1}^{n}\left(\tilde{Pr}(A_{i}(1)=1|\mathbf{X}_{i},\theta_{1}^{a(m)})-\tilde{Pr}(A_{i}(0)=1|\mathbf{X}_{i},\theta_{0}^{a(m)})\right)=\gamma(\theta_{1}^{a(m)},\theta_{0}^{a(m)}).\]

\noindent The posterior distribution of  $\gamma^{sp}$ can thus be obtained using the $M$ samples of $\theta_{0}^{a(m)},\theta_{0}^{d(m)},\theta_{1}^{a(m)}$ and $ \theta_{1}^{d(m)}$. Point and posterior interval estimates can be derived from this posterior distribution using the mean and quantiles of the distribution, respectively.

\subsection{Sensitivity Analysis} \label{Sec:Sens}

The proposed Bayesian estimation procedure relies on the strong ignorability assumption, which cannot be tested with observed data. To assess the validity this assumption we describe an interpretable sensitivity analysis.

Let $\mathbf{Z}=\{Z_1,..,Z_n\}$ denote an unobserved covariate, independent from observed covariates, with  $E[Z_i]=\mu^z_1*W_i + \mu^z_0*(1-W_i)$.  The parameters $\mu^z_1$ and $\mu^z_0$ denote the expected value of the unobserved covariate among those assigned to the active and control treatments, respectively. Without loss of generality, we set $\mu^z_1 =0$ so that $\mu^z_0$ describes the bias in $\mathbf{Z}$ between subjects assigned to receive the control intervention and subjects assigned to receive the active intervention. 
We define the following relationship between the estimated log odds of experiencing an adverse event and death, and the unobserved covariate:
\begin{align*}
\text{logit}\left(\tilde{Pr}(A_{i}(w)=1|\mathbf{X}_{i},\theta_{w}^{a})\right)=&f_{w}^{a}(g(\hat{e}(\mathbf{X}_{i})),\mathbf{\hat{B}}_{w}^{a})+\mathbf{X}_{i}^{*}\hat{\beta}_{w}^{a}+\delta_{a}\mathbf{Z}_{i}\\\text{logit}\left(\tilde{Pr}(D_{i}(w)=1|\mathbf{X}_{i},\theta_{w}^{d},A_{i}(w))\right)=&f_{w}^{d}(g(\hat{e}(\mathbf{X}_{i})),\mathbf{\hat{B}}_{w}^{d})+\mathbf{X}_{i}^{*}\hat{\beta}_{w}^{d}+\hat{\eta}_{w}A_{i}(w)+\delta_{d}\mathbf{Z}_{i}.
\end{align*}

\noindent The parameter $\delta_a$ describes the linear change in the conditional log odds of experiencing the adverse event under both treatments due to a one unit change in the unobserved covariate $\mathbf{Z}$. $\delta_d$ defines the same relationship, but for the occurrence of death. To assess the sensitivity of estimates to violations of the strong ignorability assumptions, we incorporate $\mathbf{Z}$ into the estimation procedure for different $\mu^z_0$, $\delta_a$ and $\delta_d$.

\section{Description of Data for Comparing Interventions for Type II Diabetes }  \label{DataDesc}

To illustrate the method proposed in Section \ref{sec:impute}, we analyze an observational study that compares the effects of sulfonylureas (SUs) and dipeptidyl peptidase-4 inhibitors (DPP4Is) among NH residents with type 2 diabetes mellitus (T2DM) \citep{AZThesis}. The data consists of U.S. NH residents aged 65 and older who initiated a DPP4I or SU between January 1, 2008 and September 30, 2010, and reside in a NH for at least three months preceding the initiation. After exclusions, the initial cohort included 1,064 new DPP4I users and 6,821 new SU users. The propensity score was estimated using logistic regression with a set of 198 covariates extracted from the Minimum Data Set (MDS) \citep{MDS2021} and Medicare parts A (inpatient), B (outpatient) and D (prescription drug) claims data. A one-to-one matching procedure using the propensity score yielded 1,008 patients initiating DPP4I and 1,008 initiating SU. Although propensity score matching is not necessary to apply the proposed method, we rely on the final data set from this study and estimate average effects among the treated when conducting the data analysis. This ensured that only minor imbalances among observed covariates exists between the two treatment arms. The outcomes of this study were the occurrence of adverse cardiovascular events and mortality within 180 days of initiation.

\section{Simulated Case Studies}\label{sec:casestud}

We design simulated case studies in which we control the assignment mechanism and the treatment effects for two interventions to display the differences in interpretation between estimands and to show the operating characteristics of the proposed method. The simulated data is based on the matched subjects described in Section \ref{DataDesc}. Out of the 198 covariates, we selected three continuous covariates: number of comorbidities, Morris activities of daily living scale, and number of days per week using a diuretic, and two binary covariates: an indicator of treatment for skin conditions and an indicator of hypertension. We define the probability that unit $i$ receives the active intervention as
\begin{equation} \label{simPropensity}
Pr(W_{i}=1|\mathbf{X}_{i},\alpha)=u^{-1}(\mathbf{X}_{i}\alpha),
\end{equation}

\noindent where $u$ is a link function,  $\mathbf{X}_i$ consists of the 5 covariates described above, and $\alpha$ is a coefficient vector. The expected probability that unit $i$ experiences the adverse event under treatment $w$ is 

\vspace{-0.5cm}
\begin{equation}\label{YSimfunc}
Pr(A_{i}(w)=1|\mathbf{X}_{i},\varphi{}_{w}^{a},\xi_{w}^{a})=u^{-1}(\varphi{}_{w}^{a}+\mathbf{X}_{i}\xi_{w}^{a})
\end{equation}
 
\noindent where $\varphi_w^a$ is a scalar that controls the overall probability of experiencing the adverse event outcome in treatment group $w$, and $\xi_w^a$ is a vector parameter that defines the relationship between the covariates and the adverse event outcome. The expected probability that unit $i$ experiences death under treatment $w$ is defined as

\vspace{-0.5cm}
\begin{equation} \label{ZSimfunc}
Pr(D_{i}(w)=1|\mathbf{X}_{i},\varphi{}_{w}^{d},\xi_{w}^{d},A_{i}(w))=u^{-1}(\varphi{}_{w}^{d}+\mathbf{X}_{i}\xi_{w}^{d}+\zeta_{w}*A_{i}(w))
\end{equation} 

\noindent where the parameter $\zeta_w$ defines the dependence between the adverse event and death outcomes. We examined multiple estimands to summarize the results of the data: the ITT for the adverse event and death separately, the ITT for a composite binary outcome, the SACE, $\kappa_{10}-\kappa_{01}, \frac{\kappa_{10}}{\kappa_{01}}$ and the probability that $G_{i}(w)= k$, for all $ k \in\{1,\ldots,4\}$  and $ w\in\{0,1\}$,  where $G_i(w)$ is defined:

\begin{equation} \label{G}
G_{i}(w)=
\begin{cases}
1 & \text{if patient \ensuremath{i} neither dies nor experiences heart failure under treatment \ensuremath{w}}\\
2 & \text{if patient \ensuremath{i} experiences heart failure but does not die }\\
3 & \text{if patient \ensuremath{i} does not experience the heart failure but does die}\\
4 & \text{if patient \ensuremath{i} experiences both heart failure and death.}
\end{cases}
\end{equation}

\subsection{Case Studies} \label{Case_studies}


For each case study, we assume that $u$ is the logistic function and that the values of $\xi_w^a$ are fixed at the posterior mean of the logistic regression parameters estimated using the data in Section \ref{DataDesc}. We set $\zeta_1 , \zeta_0$, $\varphi^{a}_1$, $\varphi^{a}_0 $ ,$\varphi^{d}_1$ and $\varphi^{d}_0$ to the values depicted in Appendix Tables \ref{simParams} and  \ref{simParams2}. Each setting provides different data generating mechanisms that are used to examine the operating characteristics of proposed Bayesian method. In the first case study, the adverse events and death occur together more frequently when $W_i=1$ compared to $W_i=0$. In the second case study, fewer adverse events occur when $W_i = 1$ compared to $W_i=0$, but the adverse events that occur under $W_i=1$ result in higher mortality than the adverse events that occur under $W_i = 0$.

\begin{table}[h] 
 \caption{Values of different estimands under the data generating mechanism defined in Section \ref{sec:casestud}}
    \centering
    \renewcommand{\arraystretch}{1}
\setlength{\tabcolsep}{5pt}
\begin{tabular}{c|ccc|ccc}
\hline 
\multirow{2}{*}{Estimand} & \multicolumn{3}{c|}{Case Study 1} & \multicolumn{3}{c}{Case Study 2}\tabularnewline
\cline{2-7} 
 & $w=1$ & $w=0$ & Value & $w=1$ & $w=0$ & Value\tabularnewline
\hline 
ITT Adverse & 0.380 & 0.345 & \textbf{0.035} & 0.193 & 0.226 & \textbf{-0.033}\tabularnewline
ITT Death & 0.568 & 0.538 & \textbf{0.030} & 0.378 & 0.252 & \textbf{0.127}\tabularnewline
ITT Composite & 0.634 & 0.638 & \textbf{-0.004} & 0.415 & 0.363 & \textbf{0.052}\tabularnewline
SACE & 0.152 & 0.217 & \textbf{-0.065} & 0.059 & 0.148 & \textbf{-0.089}\tabularnewline
$Pr(G_{i}(w)=1)$ & 0.366 & 0.362 & \textbf{0.005} & 0.585 & 0.637 & \textbf{-0.052}\tabularnewline
$Pr(G_{i}(w)=2)$ & 0.066 & 0.100 & \textbf{-0.034} & 0.037 & 0.111 & \textbf{-0.074}\tabularnewline
$Pr(G_{i}(w)=3)$ & 0.253 & 0.292 & \textbf{-0.039} & 0.222 & 0.136 & \textbf{0.086}\tabularnewline
$Pr(G_{i}(w)=4)$ & 0.315 & 0.245 & \textbf{0.070} & 0.156 & 0.115 & \textbf{0.041}\tabularnewline
$\kappa_{10}-\kappa_{01}$ & - & - & \textbf{0.092} & - & - & \textbf{0.120}\tabularnewline
$\frac{\kappa_{10}}{\kappa_{01}}$ & - & - & \textbf{1.445} & - & - & \textbf{1.608}\tabularnewline
\hline 
\end{tabular}
    
    \label{expectedSim1}
\end{table}

In Case Study 1, the ITT effects for the occurrence of the adverse event and mortality are 0.035 and 0.030, respectively, which implies that individuals under the active intervention suffer from more deaths and adverse effects on average. The SACE for adverse events is -0.065 which indicates that among those who survive under both treatments, less adverse events occur for those receiving the active intervention on average. The ITT effect of the composite binary outcome is -0.004. This indicates that, on average, the probability of suffering from either an adverse event or death is practically the same under both interventions. These estimand values illustrate the difficulties that can arise from using traditional estimands, in which ITT estimands show that the active intervention increases the probability of adverse events and death, but the SACE shows an effect in the opposite direction and the composite binary outcome shows no effect. 

The composite ordinal outcome provides a clearer interpretation. The probability of composite ordinal outcome levels 2 and 3 are higher under the control intervention, indicating more patients are experiencing adverse events and death separately. The probability of composite ordinal outcome level 4 is higher under the active intervention, implying that death and the adverse event are occurring more often together under the active intervention. The composite ordinal outcome estimands $\kappa_{10}-\kappa_{01}$ and $\frac{\kappa_{10}}{\kappa_{01}}$ indicate that patients are expected to do worse under the active intervention more often than they would under the control intervention. 

In Case Study 2, the ITT effects for the occurrence of the adverse event and mortality are -0.033 and 0.127, respectively. This indicates the probability of suffering from an adverse event is higher under the control intervention, however, the probability of mortality is higher under the active intervention. The SACE for the adverse events is  -0.089, which indicates that among those who survive under both treatments, the probability of experiencing an adverse event is higher for those receiving the control intervention. The ITT effect for the composite binary outcome is 0.052. This indicates that the probability of suffering from either an adverse event or death is larger under the active intervention. Each of these estimands may lead to different conclusions regarding the toxicity profiles of the interventions. The composite ordinal outcome provides a more complete description for the effects of the intervention by considering the effects on death and the adverse event simultaneously. Overall, patients are expected to do worse more often under the active intervention than they would under the control intervention. This can also be seen using $G_i(w)$, where the individuals with $W_i=1$ have higher probability of being at levels 3 and 4 of the composite ordinal outcome.

\subsection{Simulation Results}
The case studies in Section \ref{Case_studies} were replicated with 1000 datasets.  The vector parameter $\alpha$ in Equation (\ref{simPropensity}) is sampled from a multivariate normal distribution with mean 0 and identity covariance matrix at each replication. This resulted in different subpopulations assigned to the active intervention. We estimated all of the estimands in Table 1 using the proposed Bayesian Method with a natural cubic spline on the propensity score. We compared the method to a Doubly Robust (DR) estimator \citep{DR} that estimates both the propensity score and outcomes using logistic regression models with all available covariates. The SACE, $\kappa_{10}-\kappa_{01}$, and $\frac{\kappa_{10}}{\kappa_{01}}$ were not estimated using a DR method because we did not identify a valid DR method to estimate these estimands in observational studies. When defining the natural cubic spline for the proposed Bayesian method,  we start by defining six subclasses using quantiles of the propensity score distribution. In order to be able to estimate a separate cubic polynomials within a subclass, we need three observations from each treatment group. When the number of observations in each subclass for each treatment group was smaller than 3, we decrease the number of subclasses until we have at least three observations from each treatment group in each subclass. Six subclasses were selected intitially because it was shown to have good performance in many scenarios when estimating treatment effects for dichotomous outcomes \citep{RoeePaperDich}. Additionally, the Cauchy prior distribution with mode 0 and scale parameter of 2.5 has been shown to have good operating characteristics when used for Bayesian logistic regression in many scenarios, including complete separation \citep{gelman2008}. Thus, we assume independent Cauchy distributions with mode 0 and scale 2.5 as the prior distributions for all model parameters in all logistic regression outcome models.

When there is no analytical form to the posterior distribution, sampling from it can be computationally intensive when repeating for 1000 replications of each case study. To reduce the computation at each replication of each case study, we estimated the  Bayesian logistic regressions for the outcomes with the \textit{baysglm} function in the \texttt{arm} library \citep{armPackage}. This method utilizes an approximate EM algorithm to obtain the maximum a posteriori probability (MAP) estimates for the model’s parameters  \citep{gelman2008}. In large samples, the posterior distribution of these estimates can be approximated with a multivariate normal distribution centered around the MAP estimate with the expected Fisher Information as the variance-covariance matrix. In our simulations, we relied on this approximation to draw samples from the posterior distribution of $\theta_0^a$, $\theta_1^a$, $\theta_0^d$ and $\theta_1^d$. 

\begin{table}[h]
     \caption{Simulation results for the first case study}
    \centering 

\renewcommand{\arraystretch}{1}
\setlength{\tabcolsep}{1.45pt}
\begin{tabular}{ccccc|cccc}
\hline 
\multirow{2}{*}{Estimand} & \multicolumn{4}{c}{Bayesian Method} & \multicolumn{4}{c}{DR}\tabularnewline
\cline{2-9} 
 & Coverage$^{*}$ & Bias & I.W.$^{\dagger}$ & RMSE & Coverage$^{\star}$  & Bias & I.W. & RMSE\tabularnewline
\hline 
ITT Adverse & 94 & -0.001 & 0.080 & 0.021 & 95 & 0.000 & 0.083 & 0.022\tabularnewline
ITT Death & 95 & 0.000 & 0.092 & 0.023 & 96 & 0.001 & 0.097 & 0.024\tabularnewline
ITT Composite  & 94 & 0.001 & 0.085 & 0.022 & 96 & 0.001 & 0.092 & 0.023\tabularnewline
SACE & 94 & 0.001 & 0.104 & 0.028 & - & - & - & -\tabularnewline \small
$Pr(G_{i}(1)=1)-Pr(G_{i}(0)=1)$ & 94 & -0.001 & 0.085 & 0.023 & 96 & -0.001 & 0.092 & 0.023\tabularnewline \small
$Pr(G_{i}(1)=2)-Pr(G_{i}(0)=2)$ & 94 & 0.000 & 0.052 & 0.013 & 94 & 0.000 & 0.056 & 0.015\tabularnewline \small
$Pr(G_{i}(1)=3)-Pr(G_{i}(0)=3)$ & 95 & 0.002 & 0.085 & 0.022 & 96 & 0.001 & 0.092 & 0.023\tabularnewline \small
$Pr(G_{i}(1)=4)-Pr(G_{i}(0)=4)$ & 95 & -0.001 & 0.074 & 0.018 & 95 & -0.001 & 0.079 & 0.020\tabularnewline
$\kappa_{10}-\kappa_{01}$ & 96 & 0.001 & 0.102 & 0.027 & - & - & - & -\tabularnewline
$\frac{\kappa_{10}}{\kappa_{01}}$ & 96 & 0.005 & 0.399 & 0.101 & - & - & - & - \vspace{2pt}\tabularnewline
\hline 
\end{tabular}

\begin{minipage}{13.5cm}\linespread{1.0}
\footnotesize
$^{*}$ Coverage of 95\% credible interval \\
$^{\star}$ Coverage of 95\% confidence interval \\
$^\dagger$ Interval Width
\end{minipage}

     \label{fig:simres1} 
\end{table}

Table \ref{fig:simres1} depicts the performance of the estimation methods for each of the estimands in the first case study when $g$ in Equations (\ref{simPropensity}), (\ref{YSimfunc}), and  (\ref{ZSimfunc}) is the logistic link function. The Bayesian and DR methods generally result in well calibrated interval estimates. Interval lengths and root mean squared errors are slightly larger for the DR method compared to the Bayesian methods.
The proposed Bayesian method also produced well calibrated credible intervals for the SACE and for ordinal outcomes estimands. For both Bayesian and DR methods, the bias of estimates is relatively small.

\begin{table}[h]
     \caption{Simulation results for the first case study under Burr link function with $c=0.5$.}
    \centering 
\renewcommand{\arraystretch}{1}
\setlength{\tabcolsep}{2pt}
\begin{tabular}{ccccc|cccc}
\hline 
\multirow{2}{*}{Estimand} & \multicolumn{4}{c}{Bayesian Method} & \multicolumn{4}{c}{DR}\tabularnewline
\cline{2-9} 
 & Coverage & Bias & I.W. & RMSE & Coverage & Bias & I.W. & RMSE\tabularnewline
\hline 
ITT Adverse & 95 & 0.004 & 0.081 & 0.021 & 95 & 0.001 & 0.082 & 0.021\tabularnewline
ITT Death & 94 & 0.004 & 0.095 & 0.024 & 96 & 0.000 & 0.098 & 0.025\tabularnewline
ITT Composite  & 95 & 0.005 & 0.095 & 0.025 & 95 & 0.000 & 0.099 & 0.026\tabularnewline
SACE & 93 & 0.003 & 0.085 & 0.023 & - & - & - & -\tabularnewline \small
$Pr(G_{i}(1)=1)-Pr(G_{i}(0)=1)$ & 95 & -0.005 & 0.095 & 0.025 & 95 & 0.000 & 0.099 & 0.026\tabularnewline
\small $Pr(G_{i}(1)=2)-Pr(G_{i}(0)=2)$ & 94 & 0.002 & 0.060 & 0.016 & 94 & 0.000 & 0.062 & 0.017\tabularnewline
\small $Pr(G_{i}(1)=3)-Pr(G_{i}(0)=3)$ & 96 & 0.002 & 0.082 & 0.021 & 95 & -0.001 & 0.086 & 0.022\tabularnewline
\small $Pr(G_{i}(1)=4)-Pr(G_{i}(0)=4)$ & 94 & 0.002 & 0.063 & 0.017 & 95 & 0.001 & 0.067 & 0.018\tabularnewline
$\kappa_{10}-\kappa_{01}$ & 95 & 0.006 & 0.106 & 0.028 & - & - & - & -\tabularnewline
$\frac{\kappa_{10}}{\kappa_{01}}$ & 95 & 0.031 & 0.414 & 0.112 & - & - & - & - \vspace{2pt}\tabularnewline
\hline 
\end{tabular}


     \label{fig:simresBurr} 
\end{table}
In order to assess the methods' sensitivity to specification of the link function, we modify $u$ in Equations (\ref{simPropensity}), (\ref{YSimfunc}), and (\ref{ZSimfunc}) to be from the Burr family which takes the form,
\[F_c(x) = 1-(1+e^x)^{-c}.\]
When $c=1$, $F_c(x)$ corresponds to the inverse of the logistic link function. For our simulations, we set $c= 0.5$ as in Gutman and Rubin\cite{RoeePaperDich}. Table \ref{fig:simresBurr} describes the performance of the Bayesian and DR estimating methods for each of the estimands in the first case study with the Burr link function. The Bayesian estimation method results in approximately nominal coverage for all traditional and ordinal ITT estimands under the misspecified link function (Table \ref{fig:simresBurr}). The 95\% confidence intervals for the DR methods have similar coverage rates, interval length and RMSE. Similar results are observed for the second case study (Tables \ref{fig:simres2} and \ref{fig:simresBurr2} in the Appendix).

\section{Comparing Interventions  for Type II Diabetes Mellitus among Nursing Home Residents} \label{dataAnal}

We analyze the effect of treatment of T2DM with DPP4Is and SUs in NH residents using the proposed Bayesian method and the data described in Section \ref{DataDesc}. Because we use the propensity score matched sample, we estimate the treatment effects among those treated with DPP4Is.  Let $W_i = 1$ indicate treatment initiation with DPP4I and $W_i=0$ indicate treatment initiation with SU. $D_i(w)$ and $A_i(w)$ indicate the occurrence of death and heart failure, respectively, under treatment $w$ for patient $i$.  Based on the functional forms described in Equations (\ref{Yfunc}) and (\ref{Zfunc}), we assume:
\begin{align*}
A_{i}(w)&\sim\text{Bern}(\text{logit}^{-1}(f_{w}^{a}(g(\hat{e}(\mathbf{X}_{i})),\mathbf{B}_{w}^{a})+\mathbf{X}_{i}^{*}\cdot\beta_{w}^{a}))\\D_{i}(w)&\sim\text{Bern}(\text{logit}^{-1}(f_{w}^{d}(g(\hat{e}(\mathbf{X}_{i})),\mathbf{B}_{w}^{d})+\mathbf{X}_{i}^{*}\cdot\beta_{w}^{d}+A_{i}(w)\cdot\eta_{w}))\\\beta_{w}^{\ell}&\overset{i.i.d}{\sim}N\left(0,\frac{3^{2}}{\lambda_{\beta_{w}^{\ell}}}\right)\ \text{for }\ell\in\left\{ a,d\right\} ,w\in\left\{ 0,1\right\} \\\mathbf{B}_{w}^{\ell}&\overset{i.i.d}{\sim}N\left(0,8^{2}\right)\ \ \text{for }\ell\in\left\{ a,d\right\} ,w\in\left\{ 0,1\right\} \\\eta_{w}&\sim N(0,8^{2})\ \ \text{for}\ w\in\left\{ 0,1\right\} \\\lambda_{\beta_{w}^{\ell}}&\sim\text{half-Cauchy}(0,1)\ \ \text{for }\ell\in\left\{ a,d\right\} ,w\in\left\{ 0,1\right\} , 
\end{align*}

\noindent where $f(\hat{e}(\mathbf{X}_{i}),\mathbf{B}_{w}^{a})$ is a natural cubic spline on the logit of the estimated propensity score with 5 internal knots, $\mathbf{B}^a_w$ and $\mathbf{B}_{w}^{d}$ represent vectors of unknown spline coefficients, $\beta_{w}^{a}$ and $\beta_{w}^{d}$ are unknown linear adjustment coefficients, and $\eta_w$ is an unknown coefficient defining the relationship between the adverse event and mortality. Because we already include a spline on the logit of the estimated propensity score and there are a large number of covariates in each regression model, we assume  $\beta_{w}^{\ell} \overset{i.i.d}{\sim}N\left(0,\frac{3^{2}}{\lambda_{\mathbf{B}_{w}^{\ell}}}\right)$ and $\lambda_{\beta_{w}^{\ell}} \sim \text{half-Cauchy}(0,1) \text{ for }\ell\in\left\{ a,d\right\} ,w\in\left\{ 0,1\right\}$ . These prior distributions shrink the estimates of the parameters towards zero and are commonly referred to as the Ridge shrinkage prior distribution \citep{Ridge}. To test the sensitivity of the results to the prior distribution, we also conducted the analysis assuming the components of $\beta_{w}^{\ell}$ independently follow the Laplace distribution with location parameter 0 and scale parameter $\frac{3^{2}}{\lambda_{\beta_{w}^{\ell}}} \text{ for }\ell\in\left\{ a,d\right\}, w\in\left\{ 0,1\right\}$. This prior distribution is commonly referred to as the Lasso shrinkage prior distribution \citep{Ridge}. The results are quantitatively similar for both prior distributions and we report only the ones using Ridge prior distributions.

Treatment effects were summarized with both traditional estimands and composite ordinal outcome estimands using $G_i(w)$, where $G_i(w)$ is defined as it is in Equation (\ref{G}). Because we are analyzing treatment effects in the super-population, we estimated the estimands by sampling from the posterior distributions of functions of $\theta_0^a$, $\theta_1^a$, $\theta_0^d$ and $\theta_1^d$ using the sampling procedure in Section \ref{sec:impute}. To sample from the posterior distributions we used the JAGS software \citep{jags} and the rjags package \citep{runjags}. The Gelman-Rubin convergence statistics for all model parameters were less than 1.1 and posterior predictive checks demonstrated suitable model fit (see Figures \ref{GRstat} and \ref{PPcheck} in Appendix). 

\subsection{Results}
Table \ref{tab:TraditionalEstimands} summarizes the results for the ITT for heart failure and death, the composite binary outcome and the SACE. Except for the death ITT, the other three estimands are greater than zero, but do not reach the 5\% significance level. Nursing home residents with T2DM have 2\% (95\% CrI [-0.01, 0.04]) higher probability of heart failures within 180 days when treated with DPP4I (0.08, 95\% CrI [0.07, 0.09]) compared to SU (0.06, 95\% [0.05, 0.08]). Mortality within 180 days is similar between DPP4I (0.27; 95\% CrI [0.24, 0.29]) and SU (0.27; 95\% CrI [0.24, 0.29]). Lastly, among NH residents who survive for 180 days under both drugs, those who used DPP4I have 1\% (95\% CrI [0.00, 0.04]) higher probability of experiencing heart failure compared to those that use SUs.
\begin{table}[h] 
 \caption{Posterior mean and 95 \% super-population credible interval for traditional estimands estimated using the proposed Bayesian procedure}
    \centering 
        \renewcommand{\arraystretch}{1}
\setlength{\tabcolsep}{3pt}
\begin{tabular}{ccccc}
\hline 
\multirow{1}{*}{Treatment} & Heart Failure & Death & Composite Binary & SACE\tabularnewline
\hline 
DPP4I & $\underset{[0.07,\ 0.09}{0.08}$ & $\underset{[0.24,\ 0.29]}{0.27}$ & $\underset{[0.28,\ 0.33]}{0.30}$ & $\underset{[0.03,\ 0.05]}{0.04}$\tabularnewline
SU & $\underset{[0.05,\ 0.08]}{0.06}$ & $\underset{[0.24,\ 0.29]}{0.27}$ & $\underset{[0.26,\ 0.31]}{0.29}$ & $\underset{[0.02,\ 0.04]}{0.027}$\tabularnewline
\hline 
Difference & $\underset{[-0.01,\ 0.04]}{0.02}$ & $\underset{[-0.04,\ 0.04]}{0.00}$ & $\underset{[-0.01,\ 0.04]}{0.02}$ & $\underset{[0.00,\ 0.03]}{0.01}$\tabularnewline
\hline 
\end{tabular}

     \label{tab:TraditionalEstimands}

\end{table}

Table \ref{tab:OrdinalEstimandResults} presents the results for treatment effects with ordinal outcomes. After 180 days from the initiation of either DPP4I or SU for T2DM, DPP4I and SU users had similar risk of experiencing heart failure but not death (Risk
Difference (RD): 0.01, 95\% CrI: [0.00, 0.02]), death but not heart failure (RD: -0.01, 95\% CrI: [-0.04, 0.03]) and both heart failure and death (RD: 0.01, 95\% CrI: [-0.01, 0.02]). The mortality under both treatments is similar, with approximately 27\% of patients having outcomes in level 3 and 4 for both DPP4I and SU. Nursing home residents with T2DM 
were estimated to have a 5\% higher risk of a worse composite ordinal outcome under DPP4I compared to SU, but this estimate was not statistically significant at the 5\% nominal level ($\kappa_{10}/\kappa_{01}$: 1.05, 95\% CrI: [0.87, 1.24]). The estimated posterior probability that a resident selected at
random would experience a worse composite ordinal outcome under DPP4I compared to SU is 0.72.

\begin{table}[h] 
     \caption{Posterior mean and 95 \% super-population credible interval for proportion of patients who would experience each level of the outcome under each treatment}
    \centering 
        \renewcommand{\arraystretch}{1}
\setlength{\tabcolsep}{3pt}
\begin{tabular}{ccccc}
\hline 
\multirow{1}{*}{Treatment} & $Pr(G_{i}(w)=1)$ & $Pr(G_{i}(w)=2)$ & $Pr(G_{i}(w)=3)$ & $Pr(G_{i}(w)=4)$\tabularnewline
\hline 
DPP4I & $\underset{[0.68,\ 0.73]}{0.70}$ & $\underset{[0.02,\ 0.04]}{0.03}$ & $\underset{[0.19,\ 0.24]}{0.22}$ & $\underset{[0.04,\ 0.06]}{0.050}$\tabularnewline
SU & $\underset{[0.69,\ 0.74]}{0.71}$ & $\underset{[0.01,\ 0.03]}{0.02}$ & $\underset{[0.20,\ 0.25]}{0.22}$ & $\underset{[0.03,\ 0.06]}{0.04}$\tabularnewline
\hline 
Difference & $\underset{[-0.05,\ 0.03]}{-0.01}$ & $\underset{[0.00,\ 0.02]}{0.01}$ & $\underset{[0.04,\ 0.03]}{0.01}$ & $\underset{[0.01,\ 0.02]}{0.006}$\tabularnewline
\hline 
\end{tabular}

 \label{tab:OrdinalEstimandResults}
\end{table}

The third row in Table \ref{Prop_leq_table} describes the estimates of $\Delta_{j}, j\in \{1,2,3\}$ in Equation (\ref{distributionalEstimand}). At each level of the composite scale, the posterior mean of $\Delta_{j}$ is either 0 or negative. This indicates that on-average residents who received SU have lower ordinal outcome than those who receive DPP4I. However, none of these estimates are significant at the 5\% nominal level.

\begin{table}[h]
  \caption{Estimates and 95\% CI for Distributional Causal Estimands }
 \centering
         \renewcommand{\arraystretch}{1}
\setlength{\tabcolsep}{1.5pt}
\begin{tabular}{cccc}
\hline 
\multirow{1}{*}{Treatment} & $Pr(G_{i}(w)=1)$ & $Pr(G_{i}(w)\leq2)$ & $Pr(G_{i}(w)\leq3)$\tabularnewline
\hline 
DPP4I & $\underset{[0.68,\ 0.73]}{0.70}$ & $\underset{[0.71\ 0.76]}{0.73}$ & $\underset{[0.94,\ 0.96]}{0.95}$\tabularnewline
SU & $\underset{[0.69,\ 0.74]}{0.71}$ & $\underset{[0.71,\ 0.76]}{0.73}$ & $\underset{[0.94,\ 0.97]}{0.96}$\tabularnewline
\hline 
Difference & $\underset{[-0.05,\ 0.03]}{-0.01}$ & $\underset{[-0.04,\ 0.04]}{0.00}$ & $\underset{[-0.02,\ 0.01]}{-0.01}$\tabularnewline
\hline 
\end{tabular}
    \label{Prop_leq_table}
\end{table}

\subsection{Sensitivity Analysis for Data Application} \label{SensAnaly}

We use the procedure described in Section \ref{Sec:Sens} to examine the sensitivity of estimates of the relative treatment effect (RTE), $\hat{\kappa}_{10}-\hat{\kappa}_{01}$, to the strong ignorability assumption. For selected values of $\mu^z_{SU}$, we calculate $\hat{\kappa}_{10}-\hat{\kappa}_{01}$ and its standard error (SE) at different levels of $\delta_a$ and $\delta_d$. We then compute a standardized effect, $\frac{\hat{\kappa}_{10}-\hat{\kappa}_{01}}{\text{SE}(\hat{\kappa}_{10}-\hat{\kappa}_{01})}$, and plot its value at each combination of $\delta_a$ and $\delta_d$. We let the unobserved covariate $\mathbf{Z}$ be normally distributed with unit variance because covariates in our model have been standardized. Thus, $\mu^z_{SU}$ describes the standardized bias between the distribution of $Z_i$ in patients who initiated a SU and patients who initiated a DPP4I. Figure 3 displays the results for $\delta_{a},\delta_{d} \in [-1,1]$ and   $\mu^z_{SU}=1$. Under the strong ignorability assumption, $\hat{\kappa}_{10}-\hat{\kappa}_{01}$ was 0.01 (SE 0.02), which implies a standardized effect of approximately 0.5.

\begin{figure}[h]
 \centering
\includegraphics[width = 0.65\linewidth]{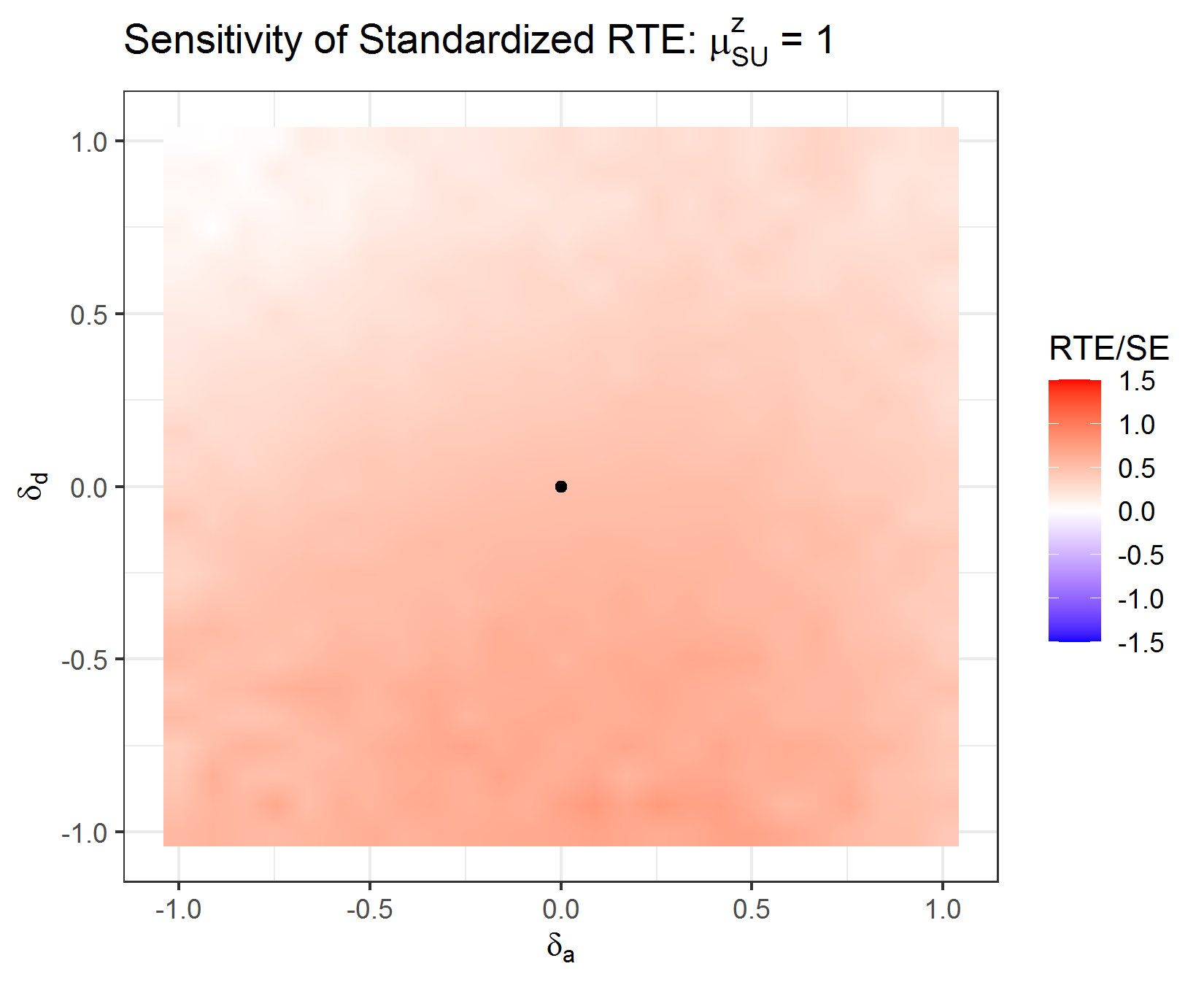} 
\vspace{-15pt}
     \caption{Sensitivity of the standardized relative treatment effect at different levels of  $\delta_a$ and $\delta_d$ when $\mu^z_{\text{SU}} =1$. The black dot denotes where $\delta_a$ and $\delta_d$ equal 0. }
    \label{Sens_0_1}
\end{figure}

\noindent All standardized values of $\hat{\kappa}_{10}-\hat{\kappa}_{01}$ are between -0.11 and  0.77, with the estimated value dropping below 0 only when $\delta_a$ approaches -1 and $\delta_d$ approaches 1. This shows that even with a relatively large standardized bias of 1 \citep{rubin2001using} the standardized estimate is between the 2.5\% to the 97.5\% percentile of the Normal distribution. These results hold even when the unobserved covariate has a conditional log odds below -5 on the adverse events and larger than 5 on death. These are large effect sizes, and none of the observed variables had a conditional log odds of this magnitude for either the adverse event or death. 

\begin{figure}[h]
 \centering
\includegraphics[width = 0.65\linewidth]{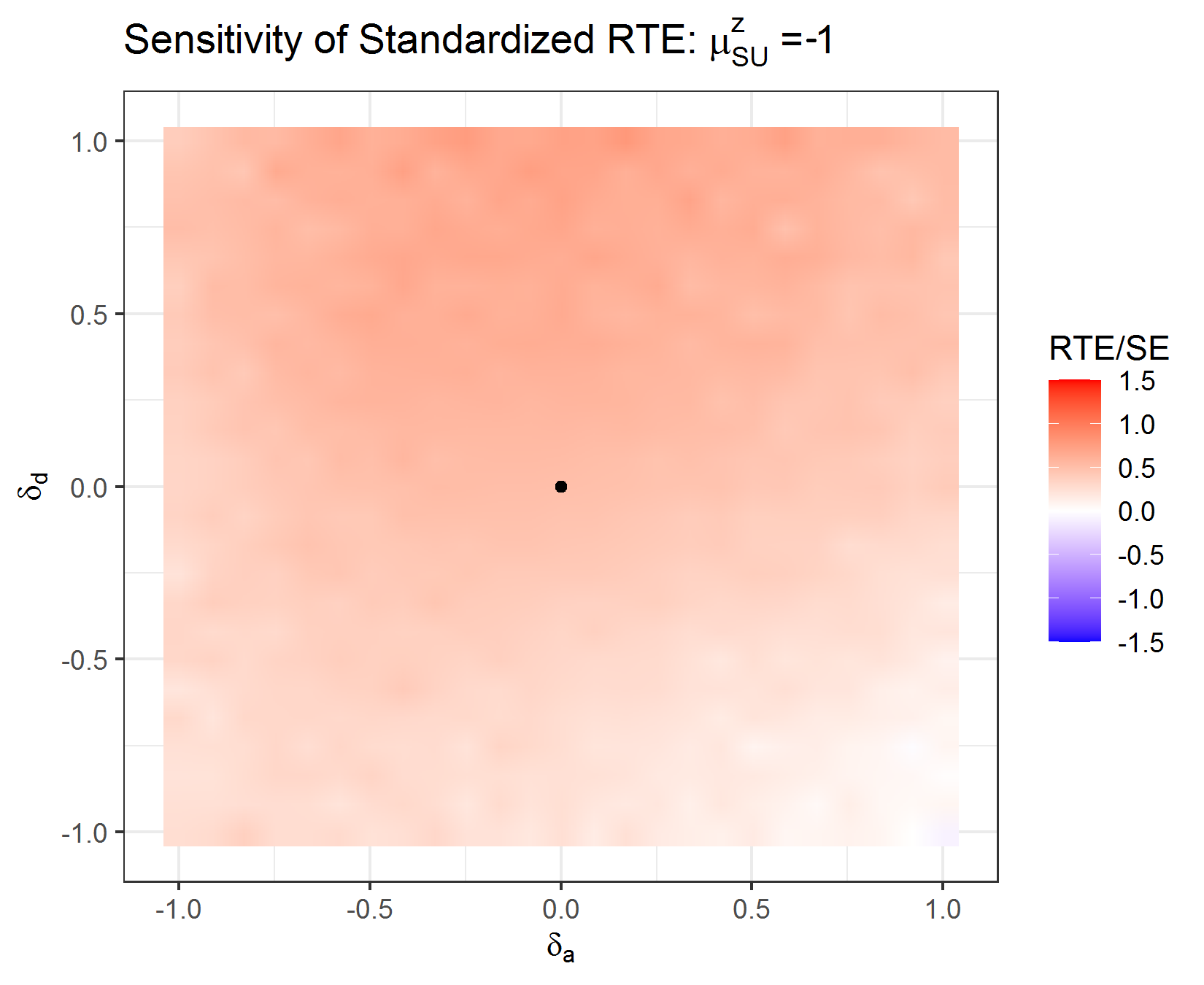} 
\vspace{-15pt}
     \caption{Sensitivity of the standardized relative treatment effect at different levels of  $\delta_a$ and $\delta_d$ when $\mu^z_{\text{SU}} =-1$. The black dot denotes where $\delta_a$ and $\delta_d$ equal 0.}
    \label{Sens_0_minus1}
\end{figure}

Figure \ref{Sens_0_minus1} depicts the sensitivity results for $\mu^z_{SU}=-1$, which displays large initial bias in the opposite direction.  
\noindent Under this configuration, the standardized effect estimates remain between -0.02 and 0.54, with estimates below 0 occurring only when $\delta_a$ approaches 1 while $\delta_d$ approaches -1. This shows that with a standardized of -1 the standardized estimates of the RTE are between the 2.5\% and the 97.5\% quantiles of the Normal distribution. Similar observations can be made with a conditional log odds of more than 5 on the adverse effects and below -5 on death.  This demonstrates that the estimates of the relative treatment effect of DPP4I compared to SU are relatively robust to violations of the strong ignorability assumption.

\section{Conclusion}

We propose a method for estimating the effects of interventions on adverse events in the presence of mortality. Our method views causal inference as a missing data problem and explicitly imputes the missing vectors of potential outcomes. This enables researchers to make inference on any finite-sample or super-population estimand. The method can be applied to observational studies with the strong ignorability assumption and to randomized trials.  For observational studies, we also provide a sensitivity analysis for the strong ignorability assumption. Combining the estimation procedure with the sensitivity analysis can generate real world evidence on possible adverse effects of interventions in the presence of differential mortality.

To address possible limitations of current estimands such as the ITT and SACE, we proposed a composite ordinal outcome to analyze the effects of interventions on adverse events. By combining the occurrence of death and an adverse event on an increasing scale of severity, the outcomes can be analyzed simultaneously so that the impact of the interventions on both is considered. The proposed method can be applied to the entire population that participated in the study, in contrast to the SACE estimand, which describes a subpopulation that may not be of interest when considering the toxicity of interventions. 

We provide case studies to display the differences in interpretation that arise from using the traditional and the composite ordinal outcome estimands. The proposed ordinal estimand can provide a more complete picture on the adverse effects of the different interventions. Simulating the case studies displays similar or superior performance of the proposed Bayesian method when compared to a Doubly Robust method for estimating  causal estimands under correctly specified and misspecified link functions.

We apply the proposed Bayesian method to an observational study that compares the effects of SUs and DPP4Is among NH residents with T2DM. Estimates of traditional estimands reveal that in this population, more heart failure is expected to occur under DPP4I than SU. Specifically, the composite ordinal outcome estimands show that  a randomly selected patient is estimated to have a 5\% higher risk of having a worse composite ordinal outcome under DPP4I than having a worse outcome under SU. However, this result was not significant at the 5\% nominal level. Prior analyses of these data utilize Cox proportional hazard models that include the occurrence of death as independent censoring when estimating the treatment effects on the rate of heart failure \cite{zullo_compareSU, AZThesis}. These models ignore the possibility that one treatment may lead to higher mortality because of more lethal heart failures or is less effective at treating T2DM. The proposed Bayesian method analyzes death and heart failure simultaneously while ranking death as the more severe adverse event to provide a more complete assessment of the risk profiles of the two interventions. This enables investigators to assess relationships between heart failure and death by producing estimates for the risk of each combination of these events under DPP4Is compared to SUs. Additionally, the proposed method provides posterior probabilties that quantify the liklihood that a patient experiences a worse composite ordinal outcome under either treatment. When using the proposed Bayesian method for this analysis, no differential risks of heart failure or death were found 180 days after initiating DPP4Is or SUs. However, at longer follow-up times, there may be benefits to using the composite ordinal outcome scale because differential risks in mortality may be more likely.

A possible limitation of the proposed composite ordinal outcome is that individuals who die during the study period may survive longer under one intervention and therefore have higher probability of experiencing an adverse event. One possible solution is to increase the number of ordinal levels in the composite outcome to distinguish between the occurrence of death during different time periods. This would require replacing the current binary model for mortality with a time-to-death model or a series of conditional binary models that represent mortality at different time points. The combination of experiencing the adverse event and death during an earlier time period would be ranked as the most severe outcome, and the combinations of later mortality and adverse events would be considered less severe outcomes. Another approach relies on a time-to-death model to make inference on the win-ratio using the imputed missing time-to-death and adverse events potential outcomes.

In conclusion, the proposed Bayesian procedure provides a new methodology to analyze observational studies with binary outcomes in the presence of mortality. The procedure imputes the adverse events and death from their joint posterior distribution which preserves the relationship between these outcomes. This relationship is used in defining a composite ordinal outcome that summarizes the effects of the interventions on the entire study population and provide a more complete picture on the toxicity profiles of these interventions. The proposed method can be used with randomized and observational studies, and for finite-sample and super-population estimands. We have applied the method for one type of adverse events, but this method can be extended to multiple types of adverse events. This extension includes incorporating additional conditional models to Equations (\ref{Yfunc}) and (\ref{Zfunc}) and adjusting the estimands accordingly. Other possible extensions may comprise models that include continuous and time-to-event outcomes. 

\ \\
\ \\
\subsubsection*{Disclosure Statement}
The authors report no conflict of interest.

\subsubsection*{Data Sharing Statement} 
The participants of this study did not give written consent for their data to be shared publicly, so due to the sensitive nature of the research supporting data is not available. However, we generate similar, synthetic, data sets for use in the supplementary materials.

\newpage

\bibliographystyle{SageV}

\newpage

\section*{Appendix}

\setcounter{table}{0}
\setcounter{figure}{0}
\setcounter{subsection}{0}
\renewcommand{\thetable}{A\arabic{table}}
\renewcommand{\thefigure}{A\arabic{figure}}
\renewcommand{\thesubsection}{A\arabic{subsection}}

\begin{table}[h] 
     \caption{Simulation Parameters for the first case study}
    \centering 
    \renewcommand{\arraystretch}{0.75}
\setlength{\tabcolsep}{2.5pt}
\begin{tabular}{ccc}
\hline 
\multicolumn{1}{c}{Parameter} & $w=1$ & $w=0$\tabularnewline
\hline 
$\varphi_{w}^{a}$ & 1 & 1\tabularnewline
$\varphi_{w}^{d}$ & 0.5 & 0.75\tabularnewline
$\zeta_{w}$ & 1.5 & 0.5\tabularnewline
$\xi_{w}^{a}$ & \{0.21, 0.32, 0.21, -2.3, -1.4\}$^{t}$ & \{0.21, 0.07, 0.21, -2.61, -2.0\} $^{t}$\tabularnewline
$\xi_{w}^{d}$ & \{-0.03, 0.18, -0.13, -1.08, -0.19\}$^{t}$ & \{0.20, 0.22, 0.08,-1.08,-0.55\} $^{t}$\tabularnewline
\hline 
\end{tabular}

    \label{simParams}
\end{table}

\begin{table}[h] 
     \caption{Simulation Parameters for the second case study}
    \centering 
    \renewcommand{\arraystretch}{0.75}
\setlength{\tabcolsep}{2.5pt}
\begin{tabular}{ccc}
\hline 
\multicolumn{1}{c}{Parameter} & $w=1$ & $w=0$\tabularnewline
\hline 
$\varphi_{w}^{a}$ & -0.25 & 0.15\tabularnewline
$\varphi_{w}^{d}$ & -0.25 & -0.75\tabularnewline
$\zeta_{w}$ & 2 & 1\tabularnewline
$\xi_{w}^{a}$ & \{0.21, 0.32, 0.21, -2.3, -1.4\}$^{t}$ & \{0.21, 0.07, 0.21, -2.61, -2.0\} $^{t}$\tabularnewline
$\xi_{w}^{d}$ & \{-0.03, 0.18, -0.13, -1.08, -0.19\}$^{t}$ & \{0.20, 0.22, 0.08,-1.08,-0.55\} $^{t}$\tabularnewline
\hline 
\end{tabular}

    \label{simParams2}
\end{table}

\begin{table}[h]
     \caption{Simulation results for the second case study.}
    \centering 

\renewcommand{\arraystretch}{1}
\setlength{\tabcolsep}{1.45pt}
\begin{tabular}{ccccc|cccc} 
\hline
\multirow{2}{*}{Estimand}         & \multicolumn{4}{c}{Bayesian Method} & \multicolumn{4}{c}{DR}             \\ 
\cline{2-9}
                                  & Coverage & Bias   & I.W.  & RMSE    & Coverage & Bias   & I.W.  & RMSE   \\ 
\hline
ITT Adverse                       & 94       & 0.001  & 0.071 & 0.019   & 94       & 0.000  & 0.072 & 0.019  \\
ITT Death                         & 95       & -0.002 & 0.086 & 0.023   & 94       & -0.001 & 0.088 & 0.023  \\
ITT Composite                     & 94       & -0.001 & 0.086 & 0.023   & 94       & -0.001 & 0.090 & 0.024  \\
SACE                              & 96       & 0.000  & 0.066 & 0.017   & -        & -      & -     & -      \\
$Pr(G_i(1)=1)-Pr(G_i(0)=1)$       & 94       & 0.001  & 0.086 & 0.023   & 95       & 0.001  & 0.090 & 0.024  \\
$Pr(G_i(1)=2)-Pr(G_i(0)=2)$       & 95       & 0.000  & 0.049 & 0.012   & 94       & 0.000  & 0.051 & 0.013  \\
$Pr(G_i(1)=3)-Pr(G_i(0)=3)$       & 94       & -0.003 & 0.074 & 0.020   & 94       & -0.001 & 0.078 & 0.021  \\
$Pr(G_i(1)=4)-Pr(G_i(0)=4)$       & 94       & 0.001  & 0.059 & 0.016   & 94       & 0.000  & 0.063 & 0.017  \\
$\kappa_{10}-\kappa_{01}$             & 95       & -0.001 & 0.093 & 0.025   & -        & -      & -     & -      \\
$\frac{\kappa_{10}}{\kappa_{01}}$ & 95       & -0.001 & 0.534 & 0.141   & -        & -      & -     & -      \\
\hline
\end{tabular}

\begin{minipage}{13.5cm}\linespread{1.0}
\footnotesize
$^{*}$ Coverage of 95\% credible interval \\
$^{\star}$ Coverage of 95\% confidence interval \\
$^\dagger$ Interval Width
\end{minipage}

     \label{fig:simres2} 
\end{table}

\begin{table}[h]
     \caption{Simulation results for the second case study under Burr link function with $c=0.5$.}
    \centering 

\renewcommand{\arraystretch}{1}
\setlength{\tabcolsep}{2pt}
\begin{tabular}{ccccc|cccc} 
\hline
\multirow{2}{*}{Estimand}         & \multicolumn{4}{c}{Bayesian Method} & \multicolumn{4}{c}{DR}             \\ 
\cline{2-9}
                                  & Coverage & Bias   & I.W.  & RMSE    & Coverage & Bias   & I.W.  & RMSE   \\ 
\hline
ITT Adverse                       & 95       & 0.004  & 0.066 & 0.017   & 95       & 0.001  & 0.064 & 0.016  \\
ITT Death                         & 94       & 0.005  & 0.080 & 0.021   & 94       & 0.000  & 0.081 & 0.022  \\
ITT Composite                     & 94       & 0.007  & 0.086 & 0.023   & 95       & -0.001 & 0.087 & 0.023  \\
SACE                              & 95       & 0.003  & 0.059 & 0.015   & -        & -      & -     & -      \\
$Pr(G_i(1)=1)-Pr(G_i(0)=1)$       & 94       & -0.007 & 0.086 & 0.023   & 95       & 0.001  & 0.087 & 0.023  \\
$Pr(G_i(1)=2)-Pr(G_i(0)=2)$       & 95       & 0.002  & 0.049 & 0.013   & 94       & 0.000  & 0.049 & 0.013  \\
$Pr(G_i(1)=3)-Pr(G_i(0)=3)$       & 94       & 0.003  & 0.070 & 0.018   & 94       & 0.000  & 0.071 & 0.019  \\
$Pr(G_i(1)=4)-Pr(G_i(0)=4)$       & 94       & 0.002  & 0.046 & 0.012   & 94       & 0.000  & 0.047 & 0.012  \\
$\kappa_{10}-\kappa_{01}$             & 94       & 0.008  & 0.091 & 0.025   & -        & -      & -     & -      \\
$\frac{\kappa_{10}}{\kappa_{01}}$ & 94       & 0.062  & 0.647 & 0.177   & -        & -      & -     & -      \\
\hline
\end{tabular}

\singlespacing 

     \label{fig:simresBurr2} 
\end{table}

\begin{figure}[h]
 \centering
\includegraphics[width =0.75\linewidth]{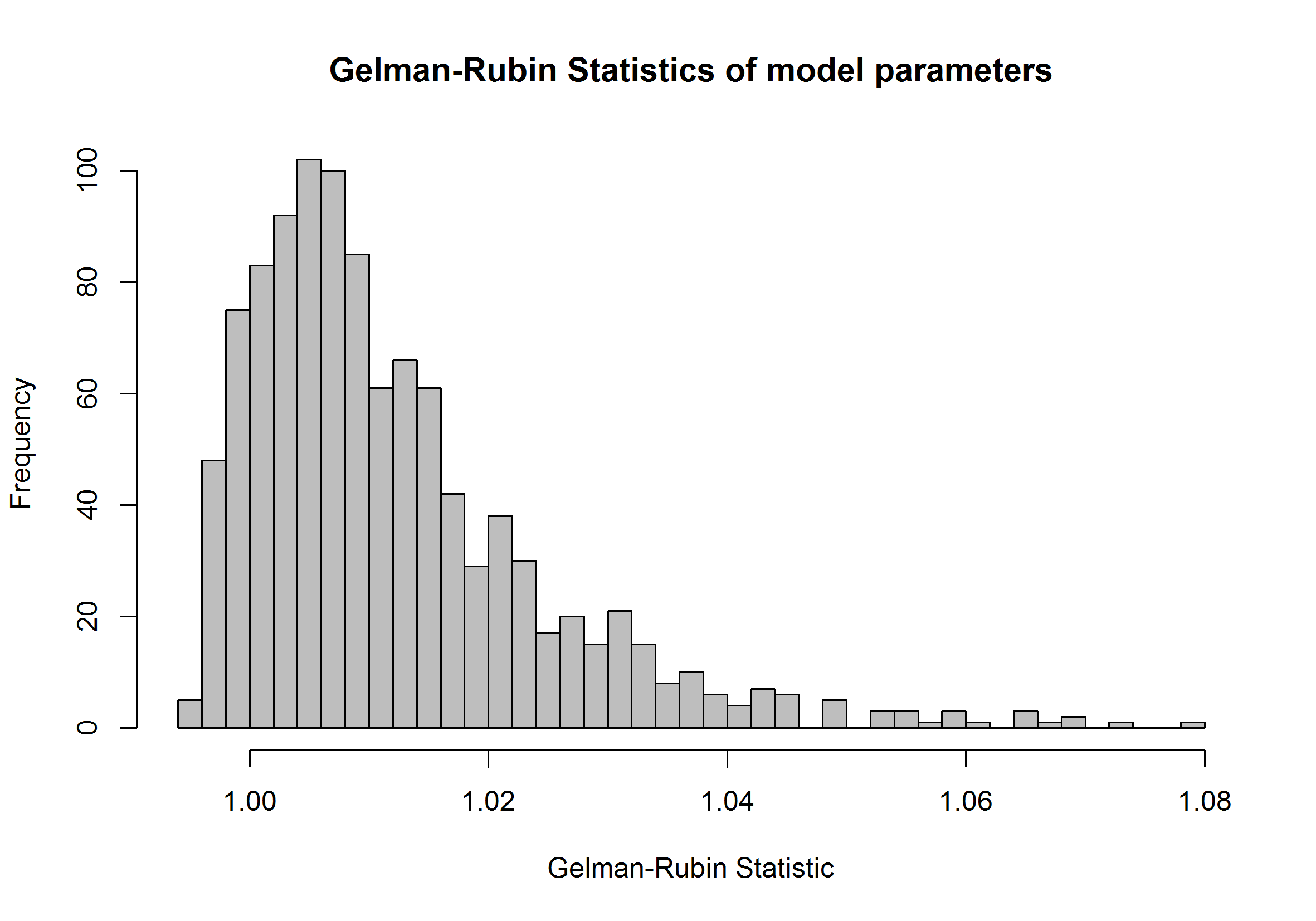} 
\vspace{-15pt}\singlespacing 
     \caption{Gelman-Rubin Statistics for model parameters used in the data analysis found in Section \ref{dataAnal}}
    \label{}
\end{figure}

\begin{figure}[h]
 \centering
\includegraphics[width =0.7\linewidth]{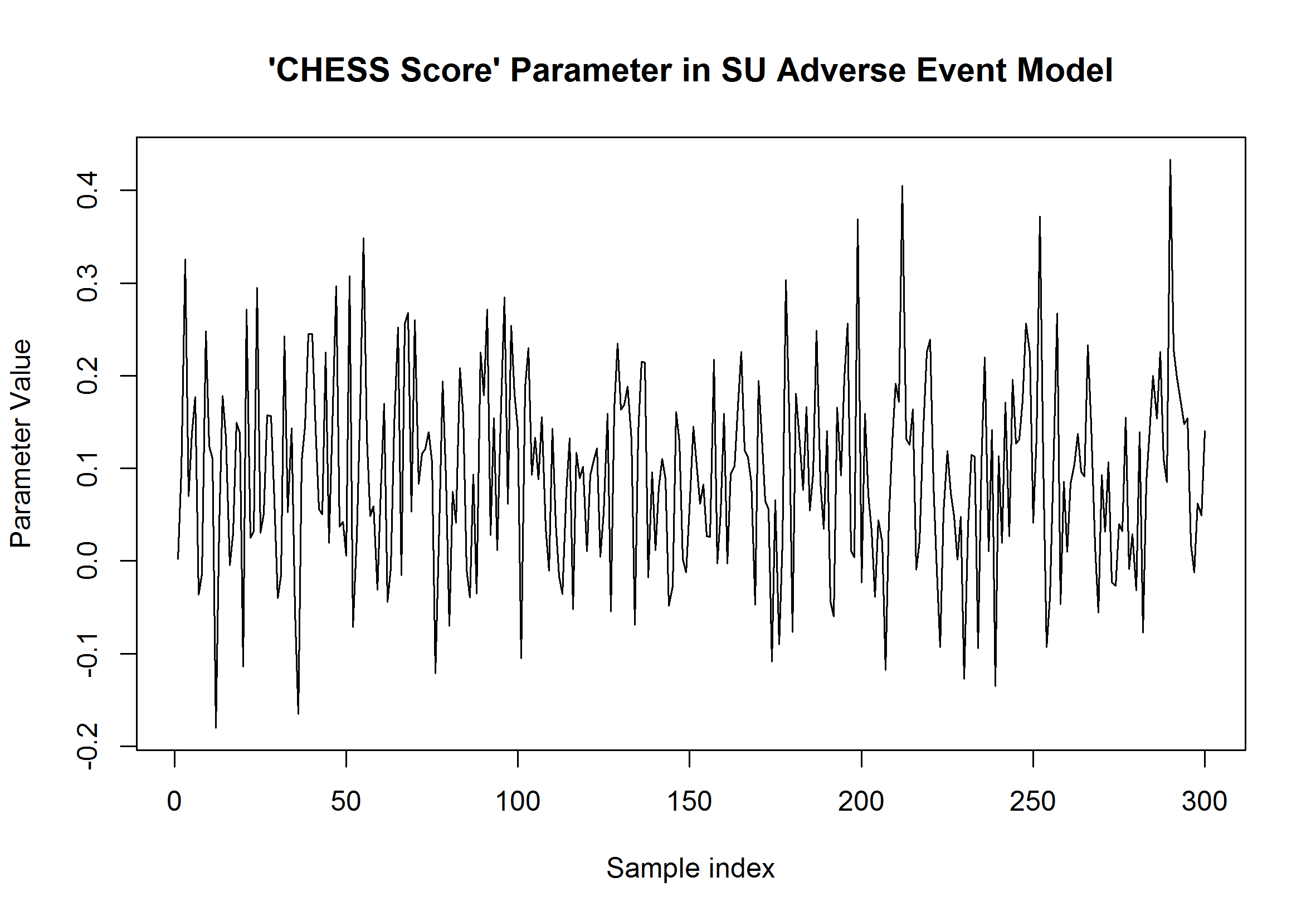} 
\vspace{-15pt}\singlespacing 
     \caption{Trace plot for the parameter related to  Changes in health, End-stage disease and Signs and
Symptoms (CHESS) score \citep{CHESS}, a health stability measure, in the logistic regression for heart failure under SU found in Section \ref{dataAnal}}

\label{GRstat}

\end{figure}

\begin{figure}[h]
 \centering
\includegraphics[width =0.7\linewidth]{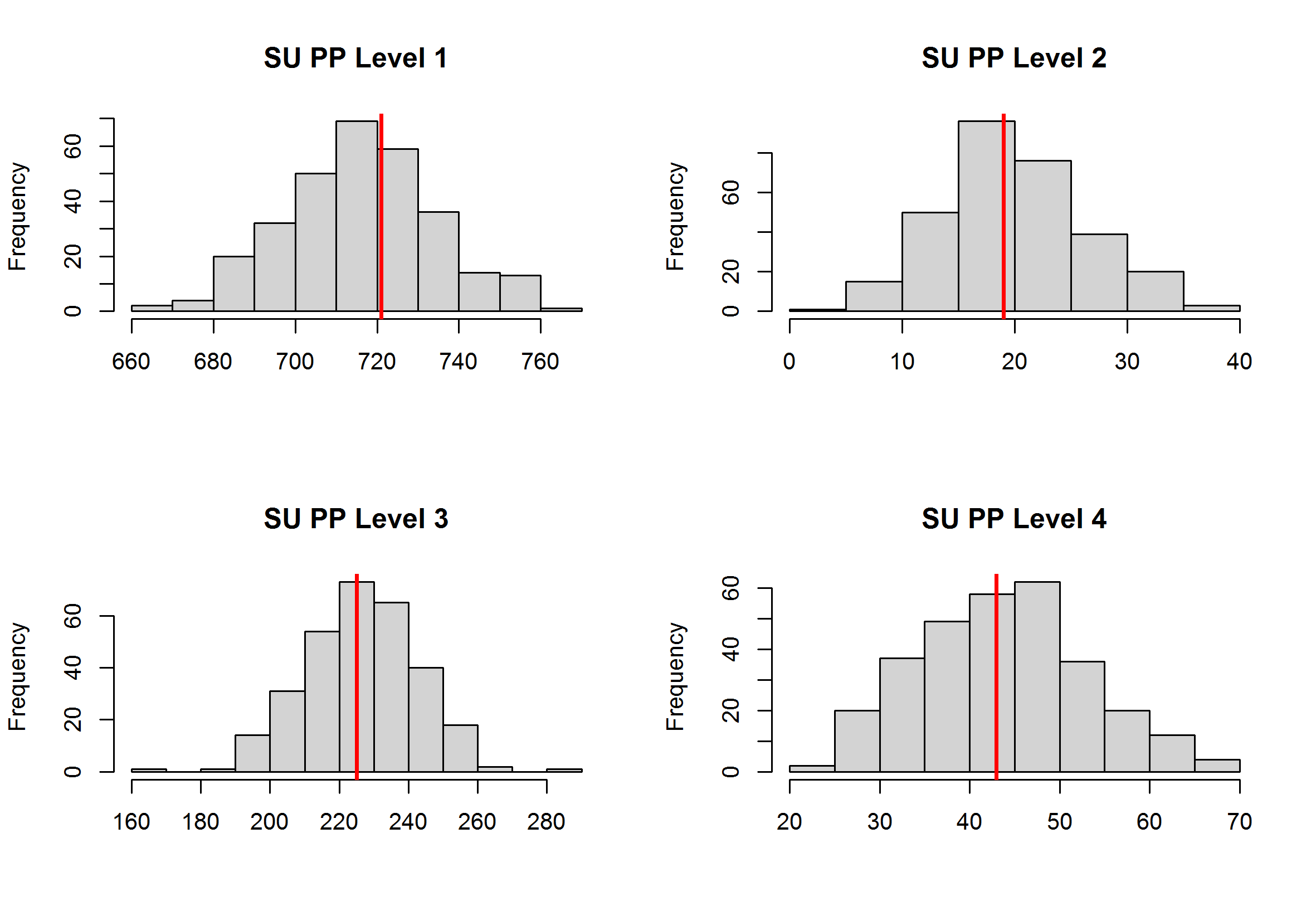} 
\vspace{-15pt}\singlespacing 
     \caption{Posterior predictive distributions for composite ordinal outcome levels under SU using the Bayesian models fit in Section \ref{dataAnal}. The vertical red line represents observed number of patients in that level of the composite ordinal outcome.}
\label{PPcheck}
\end{figure}

\newpage
\mbox{~}
\clearpage

\subsection{Multiple Imputation Combination Rules for Small Data Sets} \label{CombinationRules}

The point estimate of $\gamma$ is $\hat{\gamma}=\frac{1}{M}\sum_{i=1}^M \hat{\gamma}^{(m)}$ and the sampling variance of $\hat{\gamma}^{(m)}$ is given by $\hat{U}^{(m)}$ ($\hat{U}^{(m)}$ = 0 for finite-sample estimands). Let $\bar{U}=\frac{1}{M}\sum_{i=1}^{M}\hat{U}^{(m)}$ define the within imputation variance and let $B$ = $\frac{1}{M-1}\sum_{i=1}^{M}(\hat{\gamma}^{(m)}-\hat{\gamma})^{2}$ define the between imputation variance. The total estimate of the sampling variance for $\hat{\gamma}$ is $T= \bar{U} +(1+\frac{1}{m})B$. To obtain interval estimates in small data sets, Barnard and Rubin \cite{Tapprox} recommend approximating the distribution of $(\gamma-\hat{\gamma})T^{-1/2}$ using a $t$-distribution with $\tilde{\nu}_M$ degrees of freedom where
$$
\tilde{\nu}_M=\left(\frac{1}{\nu_M}+\frac{1}{\nu_{\widehat{obs}}}\right)^{-1}.
$$
The values of $\nu_M$ and $\nu_{\widehat{obs}}$ are given by
$$\nu_{M}=(M-1)\left(\frac{T}{\left(1+M^{-1}\right)B}\right)^2$$
and
$$ \nu_{\widehat{obs}}=\left(\frac{\nu_{\text{com}}+1}{\nu_{\text{com}}+3}\right)\nu_{\text{com}}\left(1-\frac{(1+M^{-1})B}{T}\right)$$
where $\nu_\text{com}$ is the complete-data degrees of freedom.

\newpage

\section{Supplementary Material}

Supplementary materials for this research article can be found \hyperlink{https://github.com/AnthonySisti/Adverse-Effects-Estimation-in-Observational-Studies-with-Truncation-by-Death}{on Github}. 

\begin{itemize}
    \item $\textbf{R Code for Simulation Study}$: R program that conducts the simulated case studies in Section \ref{Case_studies} of the article on synthetic data. (Simulated\_Case\_Study.R) 

    \item $\textbf{R Code for Data Analysis and Sensitivity Analysis}$:  R program that conducts a similar data analysis as the one in Sections \ref{dataAnal} and \ref{SensAnaly} of the article on synthetic data. (Simulated\_Data\_Analysis\_and\_Sensitivity.R)

    \item $\textbf{Synthetic data sets}$: Synthetic data sets for the simulation study (Synth\_CaseStudy\_X.xls) and data analysis (Synth\_Analy\_X.xls) generated using the real data described in Section \ref{DataDesc}. 
\end{itemize}

\end{document}